\documentclass[utf8]{frontiersFPHY} 
\pdfoutput=1

\usepackage{geometry,amsmath,amsfonts}
\usepackage{slashed}
\usepackage{epsfig}
\usepackage{latexsym}
\usepackage{graphicx}
\usepackage[justification=RaggedRight]{caption}
\usepackage{amssymb}
\usepackage{stfloats}
\usepackage{color}
\usepackage{multirow}
\usepackage{color}
\usepackage{rotating}
\usepackage{ifthen}
\def\firstAuthorLast{G. Arcadi}
\def\Authors{Giorgio Arcadi$^{1,2}$  }
\def\Address{\small
$^1$ Dpt. MIFT, %di Scienze Matematiche e Informatiche, Scienze Fisiche e Scienze della Terra, 
Universita degli Studi di Messina, Via Ferdinando Stagno d'Alcontres 31, I-98166 Messina, Italy.\\
$^2$ INFN Sezione di Catania, Via Santa Sofia 64, I-95123 Catania, Italy 
}
\def\corrAuthor{Giorgio Arcadi} 
\def\corrEmail{giorgio.arcadi@unime.it}

\begin{document}
\onecolumn
\firstpage{1}

\title[Non thermal DM]{
Thermal and non-thermal DM production in non-Standard Cosmologies: a mini review} 

\author[\firstAuthorLast ]{\Authors} %This field will be automatically populated
\address{} %This field will be automatically populated
\correspondance{} %This field will be automatically populated

\extraAuth{}

\maketitle

\begin{abstract}

\section{}
We provide a short review of some aspects of Dark Matter production in non Standard Cosmology. Considering the simplest Higgs portal model as definite particle physics setup, we consider the impact on the parameter space corresponding to the correct relic density, and the complementary experimental constraints, of the presence, during thermal production, of an exotic component dominating the energy density of the Universe. In the second part of the work we will focus on the case that such exotic component satisfies the equation of state of matter and it can produce DM non-thermally.
\end{abstract}

\section{Introduction}

The solution of the Dark Matter (DM) puzzle is one of the biggest challenges of Modern Particle Physics. The determination of the mechanism for DM production is a key ingredient for the solution of such puzzle. Thermal freeze-out is one of the most popular proposals as it relates the DM relic density to a single particle physics input, the so called thermally averaged DM pair annihilation cross-section. Furthermore, the value, for the latter quantity, favoured by cosmological observations of the DM abundance (see \cite{Planck:2018vyg}), is characteristic of the weak interactions, hence leading to the so-called WIMP (Weakly Interacting Massive Particles) miracle. The thermal freeze-out paradigm is however in increasing tension with null results from DM searches, especially the ones based on the principle of Direct Detection (DD) (see e.g. \cite{Arcadi:2017kky,Arcadi:2024ukq} for some reviews). In light of this, an alternative production mechanism, dubbed freeze-in \cite{Hall:2009bx} is gaining increasing attention as it can accommodate the correct relic density for very small values of the coupling between the DM and the SM states, encompassing the aforementioned experimental tensions. 
Both conventional freeze-in and freeze-out mechanisms rely on the hypothesis of a standard cosmological history of the Universe implying, in particular, that the DM is produced in a radiation dominated epoch. There are no reasons, besides minimality, to enforce a priori such assumption as we have no confirmed experimental evidences about the cosmological history prior to the Big Bang Nucleosynthesis (BBN). It is then interesting to consider the impact, on DM production, of a non-Standard Cosmological evolution of the Universe. By this we intend the possibility that at some epoch, comprised between the primordial inflation and the BBN, the energy budget of the Universe was dominated by an exotic component, i.e. different by ordinary (and dark) matter and radiation. Such exotic component impacts DM production in a twofold manner: it affects the Hubble expansion parameter and the evolution of the temperature of the Universe with time during DM thermal production; it might be itself a source of (non-thermal) production of DM.
In this work we will provide a brief review of some aspects of thermal and non-thermal production of DM in a non-standard cosmological history (for more extensive discussions the reader might refer, for example, to \cite{Kane:2015qea}). 

The paper is organized as follows. In section 2 we will describe the general set of equations describing DM production in presence of a single exotic component, to the energy budged of the Universe, characterized by an arbitrary equation of state parameter. Some analytical approximations for the solution of such equations, in the case that the new component is not a direct source of DM, will be provided as well. In section 3 a reference particle physics framework, i.e. the Higgs portal with scalar DM, will be introduced. The findings of section 2 will be applied to it. Finally, in section 4, the case of non-thermal production from an exotic matter component will be reviewed. Again, some examples of the solutions of the Boltzmann's equations will be applied to the scalar Higgs portal. The final section will be devoted to the conclusions.

\section{Boltzmann equations}

Following \cite{Arias:2019uol}, the most general set of Boltzmann equations describing the scenario under concern can be written as\footnote{The system of Boltzmann's equations can be actually written in the considered form if the following assumptions hold: i) DM self-interactions influence to a negligible extent its number density (they should ensure anyway thermalization of the DM particle with themselves). Such assumption can be easily satisfied by a suitable assignation of $\lambda_s$. ii) The DM is, during its whole production process, at least in kinetic equilibrium with the primordial plasma.}:

\begin{align}
\label{eq:Boltzmann}
    & \frac{d\rho_\phi}{dt}+3 (1+\omega)H \rho_\phi=-\Gamma_\phi \rho_\phi \nonumber\\
    & \frac{ds}{dt}+3Hs=\frac{\Gamma_\phi \rho_\phi}{T}\left(1-b_\chi\frac{E_\chi}{m_\phi}\right)+2\frac{E_\chi}{T}\langle \sigma v \rangle \left(n_\chi^2-n^2_{\chi,\rm eq}\right) \nonumber\\
    & \frac{dn_\chi}{dt}+3Hn_\chi=\frac{b_\chi}{m_\phi}\Gamma_\phi \rho_\phi-\langle \sigma v \rangle \left(n_\chi^2-n^2_{\chi, \rm eq}\right)
\end{align}
where $E_\chi^2 \simeq m_\chi^2+3 T^2$. $\rho_\phi$ represents the energy density of an exotic component $\phi$, with equation of state, relating its energy density and pressure, $p_\phi=\omega \rho_\phi$ which can dominate, during some stage, the energy density of the Universe. As further detailed in the next section, $\omega=0$, is the most popularly considered option, corresponding to a so-called 'Early Matter Domination period', due, for examples, to heavy metastable particles. On a similar footing one might consider additional radiation components, i.e. $\omega=1/3$. Other popular examples are kination ($\omega=1$) dominated \cite{Barrow:1982ei,Ford:1986sy,Visinelli:2017qga} and quintessence ($\omega=-1$) dominated Universe \cite{Salati:2002md,Profumo:2003hq}. Different values of $\omega$ can be expected by considering scalar field with suitable choices for their potentials, see e.g. \cite{Choi:1999xn,DiMarco:2018bnw}. Finally, scenarios of braneworld cosmologies \cite{Okada:2004nc} and modified gravity theories \cite{Catena:2004ba,Meehan:2014bya} can be described via the formalism illustrated below. For a more extensive review of possible origins of non-standard cosmologies we refer to \cite{Allahverdi:2020bys}.
The time evolution $\phi$ is governed by a decay rate $\Gamma_\phi$ which must ensure that it disappears before the onset of BBN. At the moment of decay the energy stored in $\phi$ is passed to the primordial plasma, increasing its entropy, and possibly to the DM. As just pointed, in light of the decay of $\phi$, the entropy density $s$ of the primordial plasma is not a conserved quantity any longer; consequently we need to consider explicitly a Boltzmann equation (alternatively one can consider, instead, an equation for the energy density of radiation, see e.g. \cite{Giudice:2000ex,Gelmini:2006pq,Gelmini:2006pw,Arcadi:2011ev}).
The last equation of (\ref{eq:Boltzmann})
tracks the time evolution of the DM number density $n_\chi$. Besides the Hubble expansion, it is governed by pair annihilation processes into SM pairs, described by the thermally averaged cross-section $\langle \sigma v \rangle$, which can be computed as shown in detail in \cite{Gondolo:1990dk} or via public available packages as micrOMEGAs \cite{Belanger:2006is,Belanger:2008sj} or DARKSUSY \cite{Gondolo:2004sc,Bringmann:2018lay},  and, possibly, by a non-thermal production term depending on $\Gamma_\phi$ and on a parameter $b_\chi$, measuring the fraction of the energy density of $\phi$ which gets converted into DM. 
Finally, $H$ is the Hubble expansion parameter:
\begin{equation}
    H^2=\frac{8\pi G}{3}\left(\rho_\phi+\rho_R+E_\chi n_\chi\right)
\end{equation}
with:
\begin{equation}
    \rho_R=\frac{\pi^2}{90}g_{\rm eff}(T)T^4
\end{equation}
Let's consider the case which $b_\chi=0$, so that $\phi$ influences DM production only indirectly by altering the expansion rate of the Universe. A recent extensive semi-analytical study of the solution of eq. \ref{eq:Boltzmann} has been presented in \cite{Arias:2019uol} and we summarize below the results.
The non-standard cosmology is parameterized mostly via three quantities:
\begin{equation}
    \omega,\,\,\, \kappa=\left. \frac{\rho_\phi}{\rho_R} \right \vert_{T=m_\chi},\,\,\,T_{\rm end}
\end{equation}
$\omega$ is the already mentioned equation of state parameter, $\kappa$ indicates at a reference temperature, the amount of energy density of $\phi$, compared to the one of radiation, and can be used to set the initial conditions for eq. (\ref{eq:Boltzmann}). $T_{\rm end}$ is finally the temperature at which the standard radiation domination era starts again after the $\phi$-dominated epoch. In the so called instantaneous decay approximation, it can be determined by the condition:
\begin{equation}
\label{eq:TR}
    \Gamma_\phi^2=H(T_{\rm end})^2=\frac{8\pi G}{3}\rho_R(T_{\rm end})\rightarrow T_{\rm end}^4=\frac{90}{\pi^2 g_{\rm eff}(T_{\rm end})}M_{\rm Pl}^2 \Gamma_{\phi}^2
\end{equation}
Such condition can be used to set $\Gamma_\phi$ as function of $T_{\rm end}$ without relying on a specific model. To avoid tension with BBN on has to require $T_{\rm end}\geq 4\,\mbox{MeV}$ \cite{Kawasaki:2000en,Hannestad:2004px,Ichikawa:2005vw,DeBernardis:2008zz}. There are three additional temperature (and hence time) scales:
\begin{itemize}
    \item the standard freeze-out temperature $T_{f.o.}$, i.e. the temperature at which DM annihilation becomes inefficient, and from which the decoupling of the DM from the primordial plasma starts, in absence of exotic components to the energy budget of the Universe. The freeze-out temperature can be determined by solving the following equation:
\begin{equation}
    x_{f.o.}=\frac{m_\chi}{T_{\rm f.o.}}=\log\left[\frac{3}{2}\sqrt{\frac{5}{\pi^5 g_{\rm eff}}}g_\chi m_\chi M_{Pl}\langle \sigma v \rangle \sqrt{x_{f.o.}}\right]
\end{equation}
with $g_\chi$ being the internal degrees of freedom of the DM candidate. For values of $\langle \sigma v \rangle$ of the order of the thermally favoured value, $x_{\rm f.o.}\sim 20 \div 30$.
\item The temperature $T_{\rm eq}$ from which, give an initial value for $\kappa$, the $\phi$ components becomes the dominant contribution to $H$.
\item The temperature $T_c$ at which the presence of the exotic component starts altering the evolution of the plasma temperature with the scale factor.
\end{itemize}
Having in mind these relevant scales, one can achieve a semi-analytical determination of the DM abundance in some limiting regimes:
\begin{itemize}
    \item $T_{\rm eq} \ll T_{f.o.}$: DM freeze-out occurs as in the standard radiation domination scenario. The main effect from $\phi$ is represented by the entropy injection during its decay, causing a dilution of thermal abundance of the DM. In the instantaneous decay approximation the DM relic density can be written as:
\begin{align}
    & Y_{\chi}=\frac{Y_{\chi}^T}{D}\simeq \frac{15}{2\pi \sqrt{10 g_{\rm eff}}}\frac{x_{f.o.}}{m_\chi M_{Pl} \langle \sigma v \rangle}{\left[\frac{1}{\kappa}{\left(\frac{T_{\rm end}}{m_\chi}\right)}^{1-3\omega}\right]}^{\frac{1}{1+\omega}}\,\,\,\,\,\,\omega \neq -1\nonumber\\
    & Y_{\chi}=\frac{Y_{\chi}^T}{D}\simeq \frac{15}{2\pi \sqrt{10 g_{\rm eff}}}\frac{x_{f.o.}}{m_\chi M_{Pl} \langle \sigma v \rangle}\left[1-\kappa {\left(\frac{m_\chi}{T_{\rm end}}\right)}^4\right]^{3/4}\,\,\,\,\,\,\,\omega=-1
\end{align}

\item $T_c \ll T_{f.o.} \ll T_{\rm eq}$: In such a regime, freeze-out occurs when Hubble expansion parameter is dominated by the $\phi$ component:
\begin{equation}
    H \simeq \frac{\sqrt{\rho_\phi}}{3 M_{Pl}^2}=\frac{\pi}{3}\sqrt{\frac{g_{\rm eff}}{10}}\frac{m_\chi^2}{M_{\rm Pl}}\sqrt{\frac{\kappa}{x^{3 (1+\omega)}}}
\end{equation}
We are however far enough from its decay time so that the relation between the temperature and the scale factor is the same as in standard cosmology. In such a case the DM abundance is again given by the ration of a thermal abundance and the same dilution factor defined in the previous case. However the thermal abundance differs from the standard computation as consequence of a different freeze-out time:
\begin{align}
    & Y_\chi=D^{-1} \frac{45}{4\pi}\frac{1-\omega}{m_\chi M_{\rm Pl}\langle \sigma v \rangle}\sqrt{\frac{\kappa}{10 g_{\rm eff}}}\tilde{x}_{f.o.}^{\frac{3}{2}(1-\omega)}\,\,\,\,\,\,\omega \neq 1\nonumber\\
    & Y_\chi=D^{-1}\frac{15}{2\pi}\frac{1}{m_\chi M_{Pl}\langle \sigma v \rangle}\sqrt{\frac{\kappa}{10 g_{\rm eff}}}{\left[\log\frac{x_{\rm end}}{\tilde{x}_{f.o.}}\right]}^{-1}\,\,\,\,\,\,\omega=1
\end{align}
where $x_{\rm end}=m_\chi/T_{\rm end}$ while $\tilde{x}_{f.o.}$ refers to a modified freeze-out time, with respect to a standard cosmological history, given by: 
\begin{equation}
    \tilde{x}_{\rm f.o.}=\log \left[\frac{3}{2}\sqrt{\frac{5}{\pi^5 g_{\rm eff}}}g_\chi \frac{m_{\chi}M_{Pl}\langle \sigma v \rangle}{\sqrt{\kappa}}\tilde{x}_{\rm f.o.}^{3/2 \omega} \right]
\end{equation}
\item $T_{\rm end} \ll T_{f.o} \ll T_c$. In such a case, DM freeze-out is affected by the different relation between the temperature and the scale parameter. The relic density can be approximated, this time, as:
\begin{align}
    & Y_\chi=\frac{45(1-\omega)}{4\pi}\sqrt{\frac{1}{10 g_{\rm eff}}}\frac{1}{M_{\rm Pl}\langle \sigma v \rangle}\left[\overline{T}_{\rm f.o.}^{4(\omega-1)}T_{\rm end}^{3-5\omega}\right]^{1/(1+\omega)}\,\,\,\,\,\,\omega \neq 1\nonumber\\
    & Y_{\chi}=\frac{45}{8\pi}\sqrt{\frac{1}{10g_{\rm eff}}}\frac{1}{T_{\rm end}M_{\rm Pl}\langle \sigma v \rangle}{\left(\log \frac{\overline{T}_{\rm f.o.}}{T_{\rm end}}\right)}^{-1}\,\,\,\,\,\,\omega=1
\end{align}
where the freeze-out temperature is obtained by solving the following equation:
\begin{equation}
    \overline{x}_{\rm f.o.}=\log\left[\frac{3}{2}\sqrt{\frac{5}{\pi^5 g_{\rm eff}}}g_\chi \frac{M_{\rm Pl}\langle \sigma v \rangle T_{\rm end}^2}{m_\chi}\overline{x}_{\rm f.o.}^{5/2}\right]
\end{equation}
\item $T_{f.o.} \ll T_{\rm end}$: the presence of an epoch dominated by the exotic component as no impact on DM production. The relic density is determined as in the Standard Cosmological model.
\end{itemize}

\section{Results in a specific case of study}

\begin{figure}
    \centering
    \includegraphics[width=0.3\linewidth]{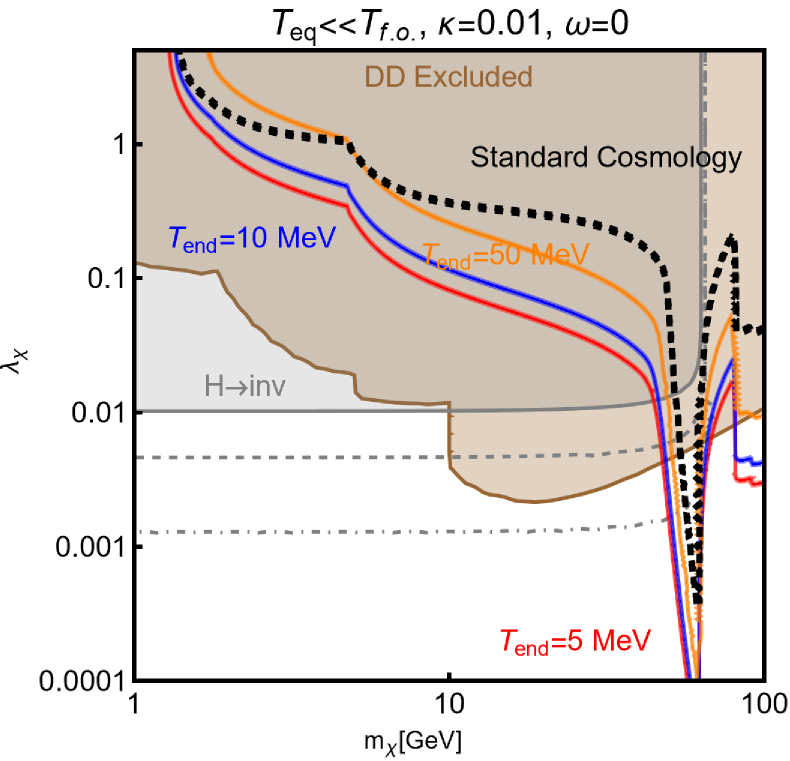}
    \includegraphics[width=0.3\linewidth]{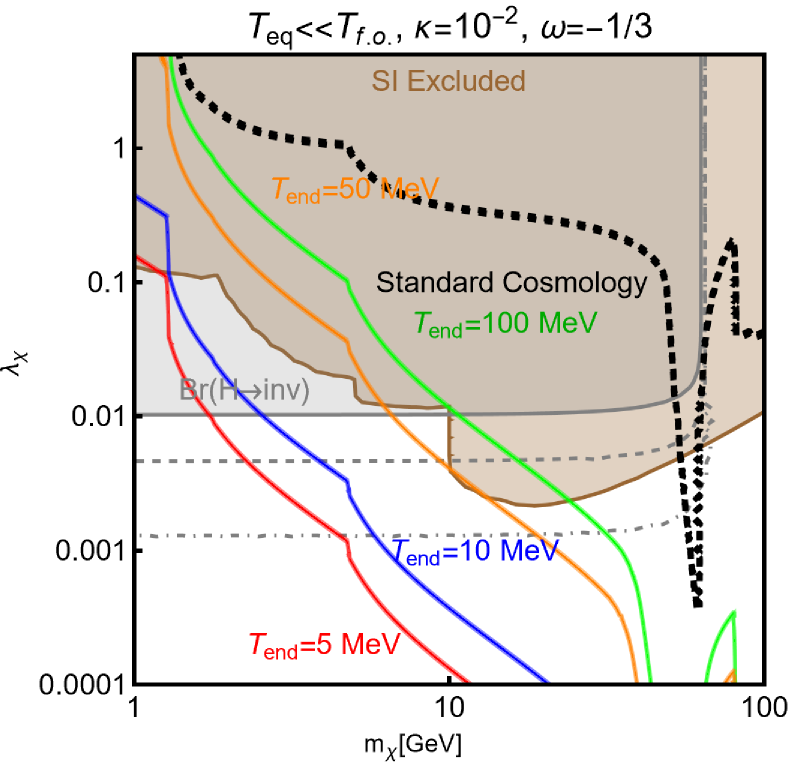}
    \includegraphics[width=0.3\linewidth]{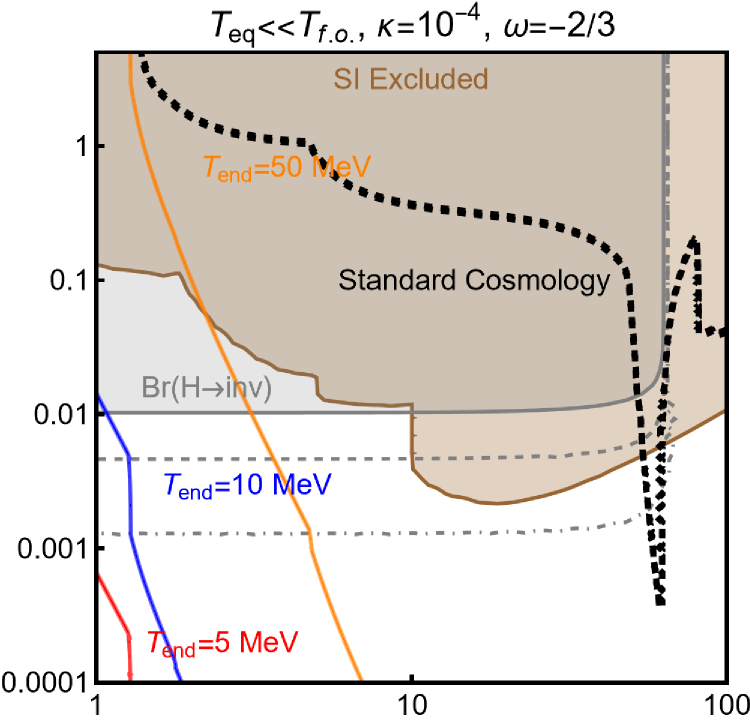}\\
    \includegraphics[width=0.3\linewidth]{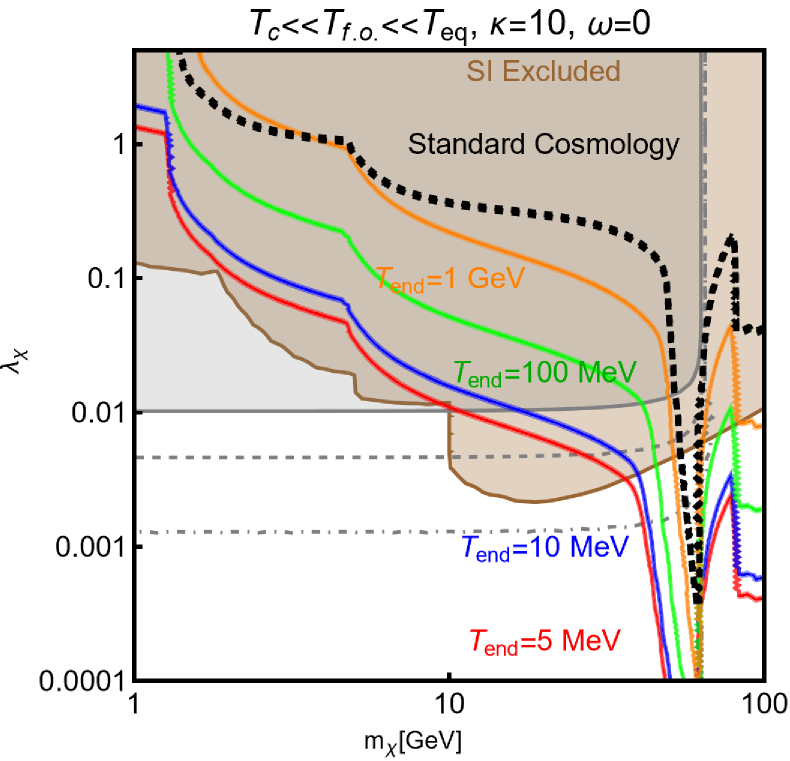}
    \includegraphics[width=0.3\linewidth]{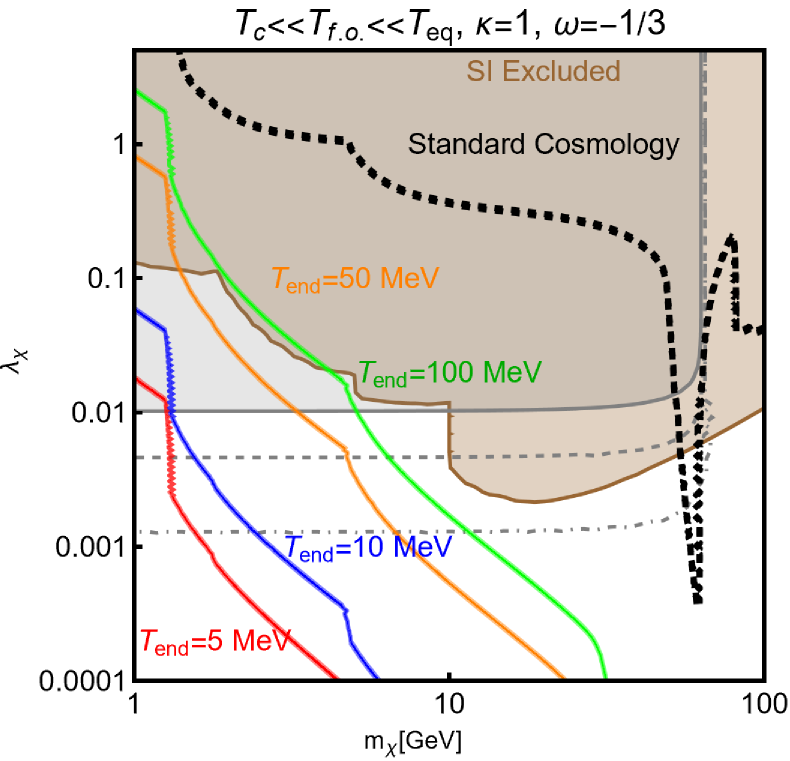}
    \includegraphics[width=0.3\linewidth]{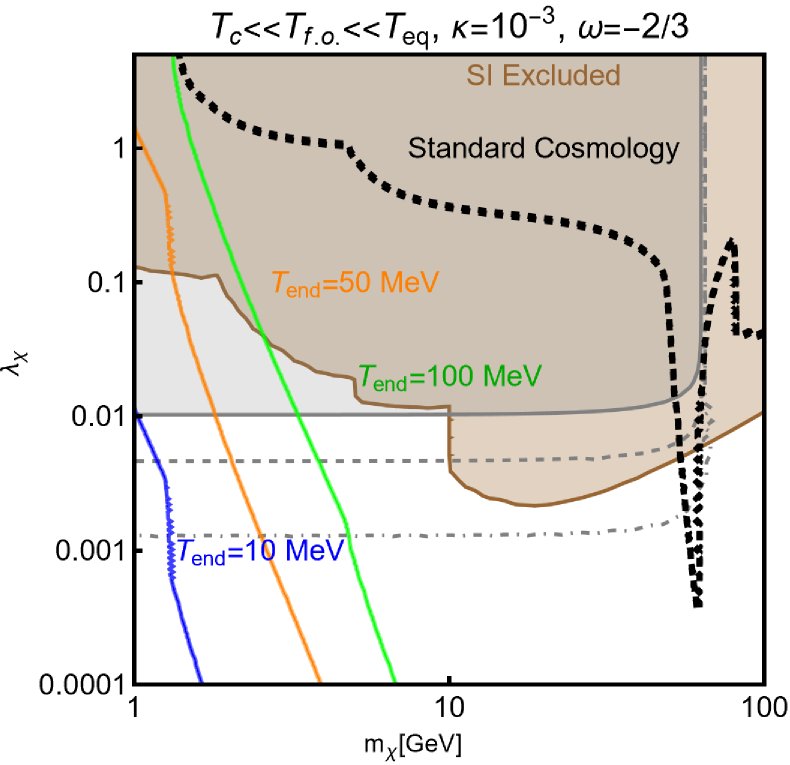}
    \caption{Parameter space of the Higgs portal considering three non-standard cosmological scenarios identified by different assignations of the $(\omega,\kappa)$ pair. The different coloured contours correspond to the correct value of the DM relic density for the assignations of $T_{\rm end}$ reported in the plots. The brown region is excluded by current DD constraints. The gray regions are, instead, ruled out by searches for invisible decays of the Higgs. The dashed gray line corresponds to future hypothetical limits $Br(H\rightarrow \mbox{inv})<0.05,0.01$. For reference we have shown, as black dashed lines, the result relative to the correct relic density in a standard cosmological scenario. The upper row refers to solution in the regime $T_{\rm end} \ll T_{\rm f.o.}$ while the bottom row to $T_{c} \ll T_{\rm f.o.} \ll T_{\rm eq}$ (see main text for discussion).}
    \label{fig:Hp1}
\end{figure}

As evident from the previous discussion, the framework under concern can be analysed in terms of a limited set of parameters without relying on a specific particle physics framework: the initial ratio $\kappa$ between the $\phi$ and the radiation energy densities, the equation of state parameter $\omega$, $T_{\rm end}$, the DM mass $m_\chi$ and the annihilation cross-section $\langle \sigma v \rangle$. It is nevertheless useful, for a better understanding of the impact of non-standard cosmologies on DM production, to consider a definite particle model. Our choice falls on the Higgs portal (see \cite{Arcadi:2019lka} for a review) with scalar DM, as it allows to maintain a low number of free parameters. Indeed, the latter model is fully characterized by the following Lagrangian:
\begin{equation}
    \mathcal{L}=-\frac{1}{2}{m_\chi^0}^2\chi^2-\frac{1}{4}\lambda_s \chi^2-\frac{1}{4}\lambda_{\chi}\chi^2 H^\dagger H
\end{equation}
with $\chi$ being a real scalar DM candidate \footnote{The phenomenology would be totally analogous in the case of complex scalar DM.} while $H$ is the Higgs doublet. After that the Higgs get vev the lagrangian generates a trilinear interaction between the Higgs boson and a pair of DM particles which allows for DM annihilations into SM fermion, gauge and Higgs boson pairs. Besides the DM mass:
\begin{equation}
m_\chi^2= {m_\chi^0}^2+\frac{1}{4}\lambda_\chi^2 v^2,   
\end{equation}
the coupling $\lambda_\chi$ is the only free parameter of the theory
By comparing the DM annihilation rate with the Hubble expansion rate of a radiation dominated era one finds that for $\lambda_\chi \gtrsim 10^{-5}$ the DM was capable of being in thermal equilibrium in the Early Universe. One could then apply the standard freeze-out paradigm and find that the correct relic density, $\Omega_\chi h^2 \approx 0.12$, is matched for $O(0.1-1)$ values of the $\lambda_\chi$ coupling, with the exception of $m_\chi \sim m_H/2$, where the s-channel resonant enhancement of the DM annihilation cross-section allows for very small values of the couplings (see also fig. \ref{fig:Hp1}). For $\lambda_\chi \leq 10^{-6}$ the DM was not capable of thermalizing with the primordial plasma. Nevertheless the correct relic density can be achieved, for $\lambda_{\chi} \sim O(10^{-11})$ via the freeze-in mechanism\footnote{Assuming that an initial DM population was produced during the inflation, the correct relic density would be achieved also for $\lambda_{\chi}=0$\cite{Arcadi:2019oxh}.}.
The parameter space corresponding to thermal freeze-out can be effectively probed experimentally. The most relevant constraints come from DM Direct Detection (DD). For our analysis we have considered the combination of the limits given by XENON1T \cite{XENON:2019gfn}, for $m_\chi \leq 10\,\mbox{GeV}$ and by LZ \cite{LZ:2022lsv} for $m_\chi \geq 10,\mbox{GeV}$. For $m_\chi \leq m_H/2$, DD constraints are well complemented by the bounds from searches of invisible decays of the SM Higgs (see e.g.~\cite{ATLAS:2023tkt} for most recent results). In this work we have adopted the limit $Br(H\rightarrow \mbox{inv})\leq 0.11$ and considered as well projected increased sensitivities to 0.05 and 0.01 \cite{Cepeda:2019klc,deBlas:2019rxi}.

Fig. \ref{fig:Hp1} illustrates, via some examples, the impact of a non-standard cosmological history, as illustrated in the previous section, on the parameter space of the scalar Higgs portal, with focus on the $m_\chi \leq 100\,\mbox{GeV}$ region, being the one most subject to experimental constraints. The different panels of the figure consider some assignations of the $(\omega,\kappa)$ pair and show, in the $(m_\chi,\lambda_\chi)$ bidimensional plane, isocontours of the correct relic density for different values of $T_{\rm end}$ ranging between 5 MeV (approximately the lower bound from BBN) to 1 GeV. To be viable, such isocontours should be (at least partially) outside the brown and gray coloured regions corresponding to, respectively, the exclusion bounds from DD and invisible Higgs decays. For reference, the isocontour corresponding to the standard freeze-out scenario has been shown as well. 
From the outcome of the plots one notices that non standard cosmologies sensitively affect the parameter space corresponding to the correct DM relic density, allowing lower values of the DM couplings and, more interestingly, lower values of the mass. In the case of an additional matter component $\omega=0$ it remains very difficult to overcome experimental constraints for $m_\chi \leq m_{H}/2$. To achieve viable parameter space a low DM masses one need to rely on more exotic components with $\omega=-1/3$ and $\omega=-2/3$.

\section{Thermal and non Thermal DM in Universe with Early Matter Domination}

The most commonly considered scenario with $b_\chi \neq 0$ is the one in which $\phi$ is an additional matter component. i.e. $\omega=0$. In this setup, $\phi$ can interpreted as a particle field always thermally decoupled from the primordial plasma. The existence of these fields is motivated in several particle physics frameworks, see e.g. \cite{Moroi:1999zb,Allahverdi:2013tca,Acharya:2008bk,Acharya:2009zt,Moroi:2013sla,Aparicio:2016qqb,Chowdhury:2018tzw} for some examples. On more recent times, primordial black holes have been proposed as this exotic matter component \cite{RiajulHaque:2023cqe}. In this kind of setups, $T_{\rm end}$ is customarily referred as reheating temperature $T_R$. While we will adopt the phenomenological determination given by eq. (\ref{eq:TR}) and use, as well, the parameter $\kappa$ to fix the initial conditions for the Boltzmann equations, one can determine $\Gamma_\phi$ from the model parameters as:
\begin{equation}
    \Gamma_\phi=D_\phi \frac{m_\phi^3}{M_{\rm Pl}^2}
\end{equation}
with $D_\phi$ depending on the specific underlying model and assign the an initial energy density $\rho_{\phi,I}=\frac{1}{2}m_\phi^2 M_{\rm Pl}^2$. In this setup, $T_R$ can be extrapolated from the numerical solution of the Boltzmann equations (see e.g. \cite{Arcadi:2011ev} for a discussion). 
Eq. (\ref{eq:Boltzmann}) can be solved, in the case of a new matter field $\Phi$, via the following change of variables \cite{Giudice:2000ex,Arcadi:2011ev}:
\begin{equation}
    \Phi=\frac{\rho_\phi a^{3}}{\Lambda},\,\,\,\,\,N_{\chi}=n_{\chi} a^3,\,\,\,\,a=\frac{A}{a_I}
\end{equation}
Such change of variables allow to gauge out the terms linear with the Hubble expansion rate, so that the system of equations gets rewritten as:

\begin{align}
    & \frac{d\Phi}{dA}=-\frac{\Gamma_\Phi}{\mathcal{H}}A^{1/2}a_I^{3/2}\Phi \nonumber\\
    & \frac{dN}{dA}=\frac{A^{1/2}a_I^{ 3/2}}{\mathcal{H}} \Lambda \frac{b_\chi}{m_\phi}\Gamma_\phi \Phi-\frac{\langle \sigma v \rangle}{\mathcal{H}}A^{-5/2}a_I^{-3/2}\left(N_\chi^2-N^2_{\chi,\rm eq}\right)\nonumber\\
    & \frac{dT}{dA}={\left(3+T\frac{dh_{\rm eff}}{dT}\right)}^{-1}\left \{- \frac{T}{A}+\frac{\Gamma_\phi \Lambda}{m_\phi}\left(1-\frac{b_\chi E_\chi}{m_\phi}\right)\frac{T}{s \mathcal{H}}A^{-5/2}a_I^{-3/2}\Phi \right. \nonumber\\
    & \left. +2 \frac{E_\chi}{s \mathcal{H}}A^{-11/2}a_I^{-9/2} \langle \sigma v \rangle \left(N_\chi^2-N^2_{\chi,\rm eq}\right) \right \}
\end{align}

where $\mathcal{H}$ is defined as:
\begin{equation}
    \mathcal{H}\equiv (a_I A)^{3/2}H={\left(\frac{\Lambda \Phi+\rho_R(T)A^3 a_I^3 +E_\chi N_\chi}{3 M_{Pl}^2}\right)}
\end{equation}
In the equations above $h_{\rm eff}$ represent the entropy effective degrees of freedom.

We will present, in the following, some examples of numerical solutions of the equations above, again adopting the Higgs portal as model for DM interactions with the SM. A similar scenario has been considered as well in \cite{Hardy:2018bph}. Before doing this we briefly illustrate some approximate solutions, following the discussion of \cite{Gelmini:2006pw,Arcadi:2011ev} (detailed studies of the Boltzmann's equations for non-thermal DM production have been conducted also in \cite{Pallis:2004yy,Drees:2017iod,Drees:2018dsj}). Assuming that $T_R$ is sufficiently lower than $T_{\rm f.o.}$, so that the thermally produced component of the DM gets sufficiently diluted to account for $\Omega_\chi$ to a negligible extent, We distinguish two leading regimes for the solution of the Boltzmann's equations. In case of that interactions between the DM and the SM are substantial, non-thermal production of DM can make the DM annihilation rate, i.e. $\Gamma_{\rm ann}= \langle \sigma v \rangle n_\chi$ to become efficient again, compared to the Hubble expansion rate, at low temperatures, so that there is a compensation between non-thermal production and annihilations. This situation occurs when the number density of non-thermally produced DM exceeds the a critical value given by:
\begin{equation}
    n_\chi^c \simeq \frac{H}{\langle \sigma v \rangle}
\end{equation}
In the instantaneous decay approximation, the condition $n_\chi > n_\chi^c$ can be re-expressed as \cite{Aparicio:2016qqb}:
\begin{equation}
\label{eq:nc}
    \frac{1}{\langle \sigma v \rangle} < b_\chi \frac{\pi^2}{30}T_R^{4/3}M_{Pl}^{2/3}
\end{equation}
evidencing that, for a fixed $b_\chi$, as $T_R$ decreases, one would need stronger DM interactions (encoded in $\langle \sigma v \rangle$) to match this condition.

In this regime, also dubbed re-annihilation regime in the literature \cite{Catena:2004ba,Cheung:2010gj}, the DM relic density can be approximated analogous expression as the standard freeze-out case, but replacing the $T_{f.o.}$ with $T_R$:
\begin{equation}
\Omega_\chi^{\rm NT}h^2 \simeq \frac{T_{f.o.}}{T_{\rm R}}\Omega_\chi^{\rm T}h^2 
\end{equation}
where $\Omega_\chi^{\rm T}$ is the relic density computed according the conventional freeze-out paradigm assuming standard cosmology (we remember $\Omega_\chi^T \propto 1/\langle \sigma v \rangle$).
If the interactions between the DM and the SM states are not efficient enough a fraction, determined by the parameter $b_\chi$, of the energy initially stored in the $\Phi$ field is directly transferred into DM particles. In such a regime the DM relic density is directly proportional to $b_\chi$ and to the reheating temperature $T_R$.

\begin{align}
    & Y_\chi (T_{\rm R})=\frac{n_\chi (T_{\rm R})}{s (T_{\rm R})}\simeq \frac{b_\chi}{m_\phi}\frac{\rho_\phi (T_{\rm R})}{s(T_{\rm R})}\simeq \frac{3}{4}\frac{b_\chi}{m_\phi}T_{\rm RH}\nonumber\\
    & \rightarrow \Omega_\chi^{\rm NT}h^2 \simeq 0.2 \times 10^4 b_\chi \frac{10 \,\mbox{TeV}}{m_\phi}\frac{T_{\rm R}}{1 \,\mbox{MeV}}\frac{m_\chi}{100\,\mbox{GeV}}
\end{align}

\begin{figure}
    \centering
    \includegraphics[width=0.45\linewidth]{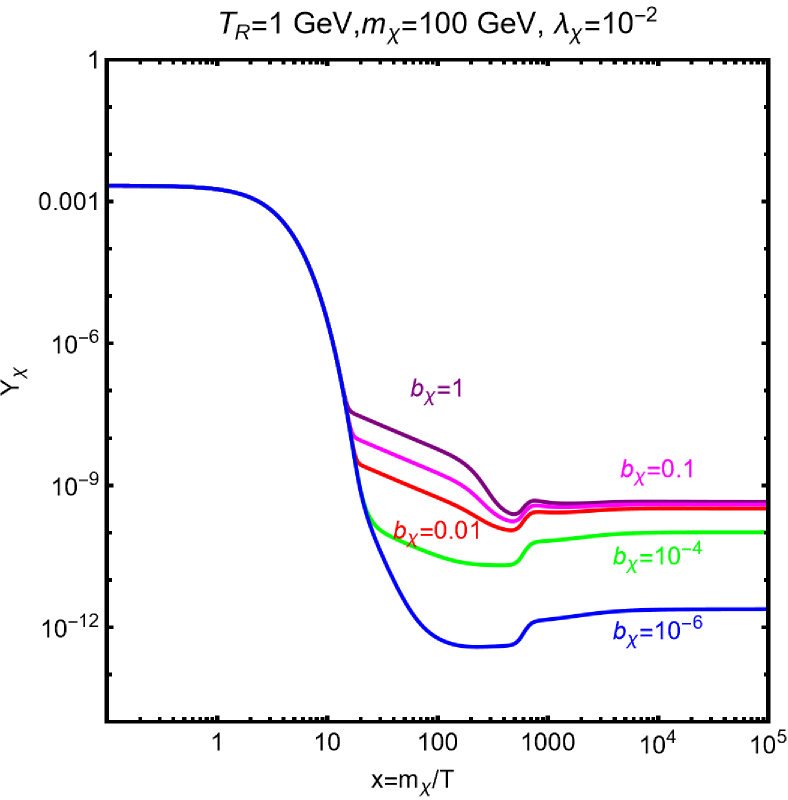}
    \includegraphics[width=0.45\linewidth]{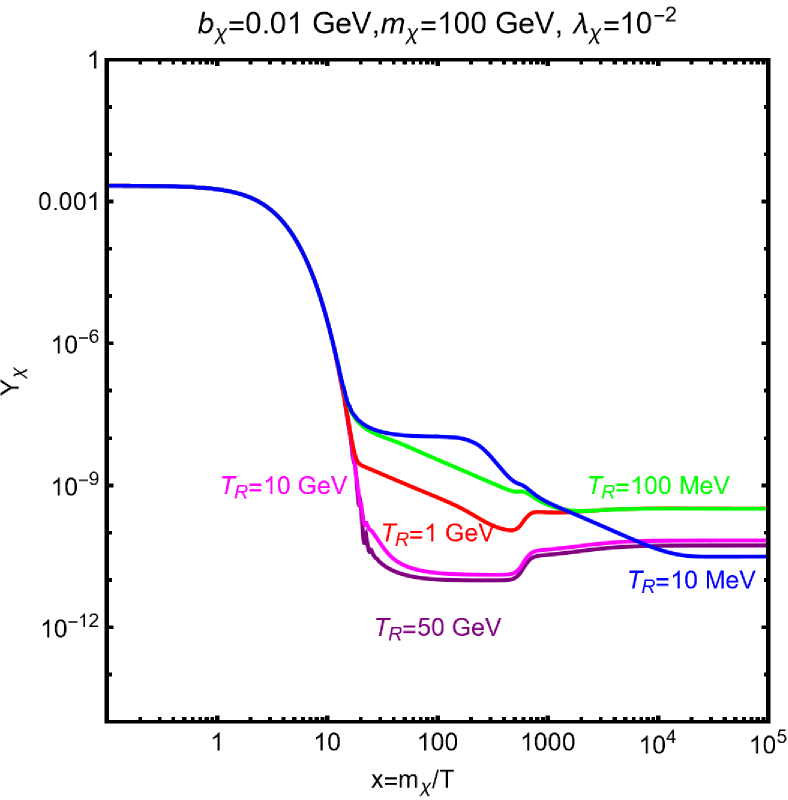}
    \caption{Evolution of the DM comoving abundance for $m_\chi=100\,\mbox{GeV}$ and $\lambda_\chi=10^{-2}$. The left plot considers the assignation $T_R=1\,\mbox{GeV}$ and different values of $b_\chi$, ranging from $10^{-6}$ to $1$, corresponding to the different colored lines. The right plots considers, instead, the fixed assignation $b_\chi=10^{-2}$ and a variation of $T_R$.}
    \label{fig:pBolpar}
\end{figure}

\begin{figure}
    \centering
    \includegraphics[width=0.5\linewidth]{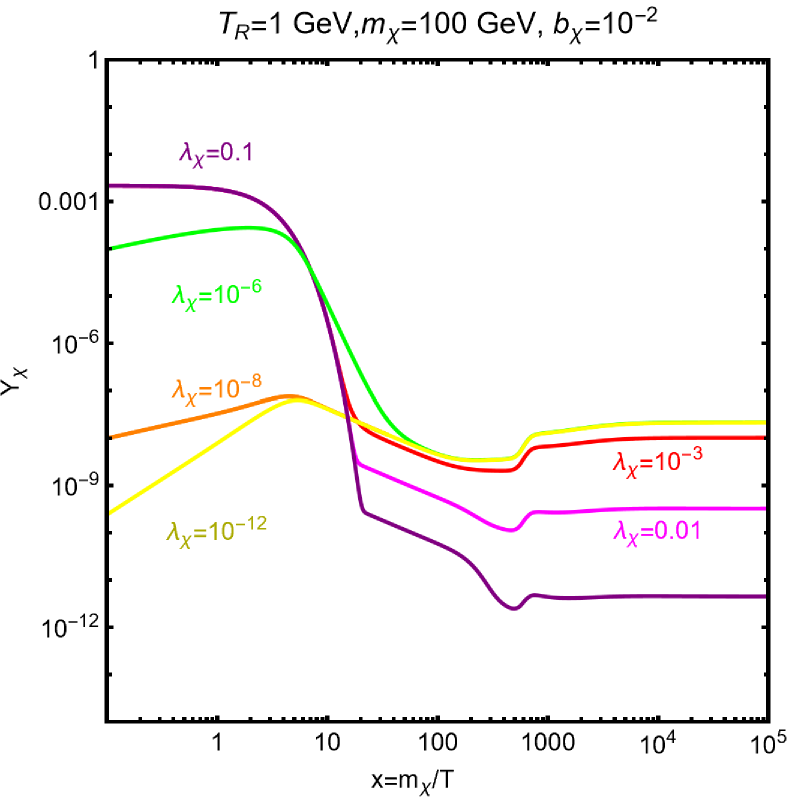}
    \caption{Evolution of the DM comoving abundance, with respect to $x=m_\chi/T$, for the assignation of the $(m_\chi,b_\chi,T_R)$ set, reported on top of the figure. The different colored lines correspond to different assignations for the DM coupling $\lambda_\chi$, reported on the plots.}
    \label{fig:pBolLambda}
\end{figure}

We illustrate in fig. \ref{fig:pBolpar} and \ref{fig:pBolLambda} some example of the solution of the Boltzmann's equations for non-thermal production of DM.

Fig. \ref{fig:pBolpar} considers the variation of the DM abundance as function of $b_\chi$ and $T_R$. In all cases $\kappa=10$ is considered. For what the particle physics input is concerned, the values of $100\,\mbox{GeV}$ and $10^{-2}$ have been considered for, respectively, the DM mass and coupling. These parameter assignations comply with the constraints from DD and Invisible Higgs decay. The left panel of fig. \ref{fig:pBolpar} considers a fixed value of the reheating temperature, namely 1 GeV, and different values of $b_\chi$. The DM abundance is very weakly dependent from the latter parameter for values above 0.01. This is because the latter is generated in the re-annihilation regime. Indeed, when abundance of the non-thermally produced DM exceeds $n_\chi^c$ (or, equivalently, $Y_\chi^c$), it is described the quasi-statical equilibrium (QSE) \cite{Cheung:2010gj} number density:
\begin{equation}
    n_\chi^{\rm QSE}={\left(\frac{b_\chi \Gamma_\phi \rho_\phi}{m_\chi \langle \sigma v \rangle}\right)}^{1/2}
\end{equation}
until the latter drops below $n_\chi^c$ and gets frozen. By further decreasing the value of $b_\chi$, the amount of non-thermally produced DM is not sufficient to reactivate annihilation processes, hence we have $Y_\chi \propto b_\chi$. 
Moving to the right panel of fig. \ref{fig:pBolpar}, we see that for $T_R < 1\,\mbox{GeV}$ increases as $T_R$ decrease, while such trend is reversed for $T_R < 100\,\mbox{GeV}$. In agreement with the discussion of ref.\cite{Gelmini:2006pq} this is due to the transition from the re-annihilation regime, corresponding to $\Omega_\chi \propto T_R^{-1}$, occurring at high reheating temperature, to the regime in which annihilations are not active, corresponding to $\Omega_\chi \propto T_R$.

In fig. \ref{fig:pBolLambda} we have considered the impact of a variation of the DM coupling $\lambda_\chi$. The values $T_R=1\,\mbox{GeV}$ and $b_\chi=0.01$ have been considered. For values of $\lambda_\chi >10^{-5}$ the DM is in thermal equilibrium in the first stages of the evolution of the Universe. For value of $x \gtrsim 10$, $Y_\chi$ starts deviating from the equilibrium value and two competing effects become relevant: the dilution by the entropy injection due to the late decay of $\Phi$ and the non-thermal production process. The final DM abundance is inversely proportional to the coupling $\lambda_\chi$ as the solution tracks the re-annihilation regime. For $\lambda_\chi < 10^{-5}$ the DM is initially produced via freeze-in. Such abundance is diluted away by the decay of the $\Phi$ field and replaced by a non-thermal population of DM without re-annihilation as the DM interaction rate is too suppressed. This explains the fact that the DM abundance is independent on the value of $\lambda_\chi$.

We conclude our analysis with few remarks about the detection prospects of the scenarios discussed in this paper. On general grounds, distinguishing non-standard cosmological scenarios only via earth-scale experiments, as the ones based on Direct/Indirect Detection as well as collider searches, is complicated as they can reconstruct DM particle properties, like the size of the interactions with SM states, while being affected to a negligible extent to the cosmological ones. It is nevertheless evident that, in presence of a non-standard cosmological history, the parameter space, corresponding to the correct relic density, can vary substantially with respect to the case of thermal freeze-out. Consequently, an hypothetical future signal, for example at a current or next generation direct detection facility, possibly incompatible with the expectations from the conventional freeze-out paradigm, would represent a very useful indication (see e.g. \cite{Arias:2019uol} for similar ideas). A conclusive statement would nevertheless require a complementary signal from a probe of pre-BBN cosmology. In this context, gravitational Wave (GW) detectors capable of probing the primordial GW background can make the difference \cite{Figueroa:2019paj,Liu:2023pau}.
As final remark we mention that scenarios of non-thermal production of DM, as the one discussed in Section 4, can, in the re-annihilation regime, already been effectively probed. Indeed Indirect Detection experiments (see e.g. \cite{Fermi-LAT:2016afa,McDaniel:2023bju}) and CMB probes (\cite{Galli:2009zc,Planck:2018vyg}) are already sensitive to DM annihilation cross-sections of the order of the thermally favoured one. Consequently, scenarios of non-thermal production in the reannihilation regime, which reproduce the correct relic density for annihilation cross-section greater than the thermal one, might be strongly constrained or ruled out. Notice anyway that this statement is strictly valid for models with s-wave dominated annihilation cross-section, i.e. the case in which the value of $\langle \sigma v \rangle$ at freeze-out and CMB/present times substantially coincide. Finally, also structure formation could provide insight about non-thermal production scenarios, as DM properties at that time could deviate from the conventional Cold Dark Matter paradigm, leaving an imprint which could be traced by Lyman-$\alpha$ \cite{Ballesteros:2020adh}.

%\section{Non-thermal production of DM}

%\section{Results}

%\begin{figure}
%    \centering
%    \includegraphics[width=0.6\linewidth]{ShportalNT1.pdf}
%    \caption{}
%    \label{fig:ShportalNT1}
%\end{figure}

%\begin{figure}
%    \centering
%    \subfloat{\includegraphics[width=0.5\linewidth]{plambda.pdf}}
%    \subfloat{\includegraphics[width=0.5\linewidth]{plambdabis.pdf}}\\
%    \subfloat{\includegraphics[width=0.5\linewidth]{plambdaG.pdf}}
%    \subfloat{\includegraphics[width=0.5\linewidth]{plambdaGbis.pdf}}
%    \caption{Caption}
%    \label{fig:my_label}
%\end{figure}

\section{Conclusions}

Thermal freeze-out is a very popular framework, leading to predictive models testable via a broad variety of complementary experimental search strategies. It relies, however on the assumption of standard cosmological history during DM production. There are no a priori reasons to enforce such assumption. We have provided a brief review of the scenario of thermal and non-thermal production of DM in a non-standard cosmological history, represented by an exotic component, possibly dominating the energy budget of the Universe during DM production and prior to BBN. Despite the result might be illustrated in a very general perspective we have found convenient to identify a reference model corresponding to the Higgs portal with scalar DM. Assuming only thermal production of DM, the non-standard cosmological evolution enlarges the parameter spaces complying with constraints from DD and invisible Higgs decays. In the second part of this review we have focused on non-thermal production focusing on the case in which the Universe encounters an early matter domination epoch. We have illustrated the relevant Boltzmann's equations and discussed both numerical and analytical approximations of the solutions.

%\appendix

%\section*{Appendix A: More general change of variables}

%\begin{equation}
%    \mathcal{H}=H a^{3 (1+\omega)/2}=\sqrt{\frac{1}{3 M_{Pl}^2}}\left(m_{DM}N a^\omega+\rho_R (T) a^{3 (1+\omega)}+\Lambda \Phi\right)
%\end{equation}

%\begin{align}
%    & \frac{d\Phi}{dA}=-\frac{\Gamma_\Phi}{\mathcal{H}}A^{\frac{1+3 \omega}{2}}a_I^{\frac{3(1+\omega)}{2}} \nonumber\\
%    & \frac{dN}{dA}=\frac{b}{m_\phi}\frac{\Lambda}{\mathcal{H}}A^{(1 +2\omega)/2}a_I^{ (3+2 \omega)/2}\Gamma_\phi \Phi-\frac{\langle \sigma v \rangle}{\mathcal{H}}A^{(3 \omega-5)/2}a_I^{3 (\omega-1)/2}\left(N^2-N^2_{\rm eq}\right)\nonumber\\
%    & \frac{dT}{dA}={\left(3+T\frac{dh_{\rm eff}}{dT}\right)}^{-1}\left \{- \frac{T}{A}+\frac{\Gamma_\phi \Lambda}{m_\phi}\left(1-b\frac{E}{m_\phi}\right)\frac{T}{s \mathcal{H}}A^{-(3 \omega+5)/2}a_I^{-3 (\omega+1)/2}\Phi \right. \nonumber\\
 %   & \left. +2 \frac{E}{s \mathcal{H}}A^{(3 \omega-11)/2}a_I^{3 (\omega -3)/2} \langle \sigma v \rangle \left(N^2-N^2_{\rm eq}\right) \right \}
%\end{align}

\bibliographystyle{frontiersinHLTH&FPHY} % for Physics and Mathematics articles
\bibliography{bibfile}

\begin{thebibliography}{56}
\expandafter\ifx\csname natexlab\endcsname\relax\def\natexlab#1{#1}\fi
\expandafter\ifx\csname urlstyle\endcsname\relax
  \expandafter\ifx\csname doi\endcsname\relax
  \def\doi#1{doi:\discretionary{}{}{}#1}\fi \else
  \expandafter\ifx\csname doi\endcsname\relax
  \def\doi{doi:\discretionary{}{}{}\begingroup \urlstyle{rm}\Url}\fi \fi
\expandafter\ifx\csname selectlanguage\endcsname\relax
  \def\selectlanguage#1{}\fi

\bibitem[{Aghanim et~al.(2020)}]{Planck:2018vyg}
Aghanim N, et~al.
\newblock {Planck 2018 results. VI. Cosmological parameters}.
\newblock {\em Astron. Astrophys.\/} {\bf 641} (2020) A6.
\newblock \doi{10.1051/0004-6361/201833910}.
\newblock [Erratum: Astron.Astrophys. 652, C4 (2021)].

\bibitem[{Arcadi et~al.(2018)Arcadi, Dutra, Ghosh, Lindner, Mambrini, Pierre et~al.}]{Arcadi:2017kky}
Arcadi G, Dutra M, Ghosh P, Lindner M, Mambrini Y, Pierre M, et~al.
\newblock {The waning of the WIMP? A review of models, searches, and constraints}.
\newblock {\em Eur. Phys. J. C\/} {\bf 78} (2018) 203.
\newblock \doi{10.1140/epjc/s10052-018-5662-y}.

\bibitem[{Arcadi et~al.(2024)Arcadi, Cabo-Almeida, Dutra, Ghosh, Lindner, Mambrini et~al.}]{Arcadi:2024ukq}
Arcadi G, Cabo-Almeida D, Dutra M, Ghosh P, Lindner M, Mambrini Y, et~al.
\newblock {The Waning of the WIMP: Endgame?}  (2024).

\bibitem[{Hall et~al.(2010)Hall, Jedamzik, March-Russell, and West}]{Hall:2009bx}
Hall LJ, Jedamzik K, March-Russell J, West SM.
\newblock {Freeze-In Production of FIMP Dark Matter}.
\newblock {\em JHEP\/} {\bf 03} (2010) 080.
\newblock \doi{10.1007/JHEP03(2010)080}.

\bibitem[{Kane et~al.(2016)Kane, Kumar, Nelson, and Zheng}]{Kane:2015qea}
Kane GL, Kumar P, Nelson BD, Zheng B.
\newblock {Dark matter production mechanisms with a nonthermal cosmological history: A classification}.
\newblock {\em Phys. Rev. D\/} {\bf 93} (2016) 063527.
\newblock \doi{10.1103/PhysRevD.93.063527}.

\bibitem[{Arias et~al.(2019)Arias, Bernal, Herrera, and Maldonado}]{Arias:2019uol}
Arias P, Bernal N, Herrera A, Maldonado C.
\newblock {Reconstructing Non-standard Cosmologies with Dark Matter}.
\newblock {\em JCAP\/} {\bf 10} (2019) 047.
\newblock \doi{10.1088/1475-7516/2019/10/047}.

\bibitem[{Barrow(1982)}]{Barrow:1982ei}
Barrow JD.
\newblock {MASSIVE PARTICLES AS A PROBE OF THE EARLY UNIVERSE}.
\newblock {\em Nucl. Phys. B\/} {\bf 208} (1982) 501--508.
\newblock \doi{10.1016/0550-3213(82)90233-4}.

\bibitem[{Ford(1987)}]{Ford:1986sy}
Ford LH.
\newblock {Gravitational Particle Creation and Inflation}.
\newblock {\em Phys. Rev. D\/} {\bf 35} (1987) 2955.
\newblock \doi{10.1103/PhysRevD.35.2955}.

\bibitem[{Visinelli(2018)}]{Visinelli:2017qga}
Visinelli L.
\newblock {(Non-)thermal production of WIMPs during kination}.
\newblock {\em Symmetry\/} {\bf 10} (2018) 546.
\newblock \doi{10.3390/sym10110546}.

\bibitem[{Salati(2003)}]{Salati:2002md}
Salati P.
\newblock {Quintessence and the relic density of neutralinos}.
\newblock {\em Phys. Lett. B\/} {\bf 571} (2003) 121--131.
\newblock \doi{10.1016/j.physletb.2003.07.073}.

\bibitem[{Profumo and Ullio(2003)}]{Profumo:2003hq}
Profumo S, Ullio P.
\newblock {SUSY dark matter and quintessence}.
\newblock {\em JCAP\/} {\bf 11} (2003) 006.
\newblock \doi{10.1088/1475-7516/2003/11/006}.

\bibitem[{Choi(2000)}]{Choi:1999xn}
Choi K.
\newblock {String or M theory axion as a quintessence}.
\newblock {\em Phys. Rev. D\/} {\bf 62} (2000) 043509.
\newblock \doi{10.1103/PhysRevD.62.043509}.

\bibitem[{Di~Marco et~al.(2018)Di~Marco, Pradisi, and Cabella}]{DiMarco:2018bnw}
Di~Marco A, Pradisi G, Cabella P.
\newblock {Inflationary scale, reheating scale, and pre-BBN cosmology with scalar fields}.
\newblock {\em Phys. Rev. D\/} {\bf 98} (2018) 123511.
\newblock \doi{10.1103/PhysRevD.98.123511}.

\bibitem[{Okada and Seto(2004)}]{Okada:2004nc}
Okada N, Seto O.
\newblock {Relic density of dark matter in brane world cosmology}.
\newblock {\em Phys. Rev. D\/} {\bf 70} (2004) 083531.
\newblock \doi{10.1103/PhysRevD.70.083531}.

\bibitem[{Catena et~al.(2004)Catena, Fornengo, Masiero, Pietroni, and Rosati}]{Catena:2004ba}
Catena R, Fornengo N, Masiero A, Pietroni M, Rosati F.
\newblock {Dark matter relic abundance and scalar - tensor dark energy}.
\newblock {\em Phys. Rev. D\/} {\bf 70} (2004) 063519.
\newblock \doi{10.1103/PhysRevD.70.063519}.

\bibitem[{Meehan and Whittingham(2014)}]{Meehan:2014bya}
Meehan MT, Whittingham IB.
\newblock {Dark matter relic density in Gauss-Bonnet braneworld cosmology}.
\newblock {\em JCAP\/} {\bf 12} (2014) 034.
\newblock \doi{10.1088/1475-7516/2014/12/034}.

\bibitem[{Allahverdi et~al.(2020)}]{Allahverdi:2020bys}
Allahverdi R, et~al.
\newblock {The First Three Seconds: a Review of Possible Expansion Histories of the Early Universe}  (2020).
\newblock \doi{10.21105/astro.2006.16182}.

\bibitem[{Giudice et~al.(2001)Giudice, Kolb, and Riotto}]{Giudice:2000ex}
Giudice GF, Kolb EW, Riotto A.
\newblock {Largest temperature of the radiation era and its cosmological implications}.
\newblock {\em Phys. Rev. D\/} {\bf 64} (2001) 023508.
\newblock \doi{10.1103/PhysRevD.64.023508}.

\bibitem[{Gelmini et~al.(2006)Gelmini, Gondolo, Soldatenko, and Yaguna}]{Gelmini:2006pq}
Gelmini G, Gondolo P, Soldatenko A, Yaguna CE.
\newblock {The Effect of a late decaying scalar on the neutralino relic density}.
\newblock {\em Phys. Rev. D\/} {\bf 74} (2006) 083514.
\newblock \doi{10.1103/PhysRevD.74.083514}.

\bibitem[{Gelmini and Gondolo(2006)}]{Gelmini:2006pw}
Gelmini GB, Gondolo P.
\newblock {Neutralino with the right cold dark matter abundance in (almost) any supersymmetric model}.
\newblock {\em Phys. Rev. D\/} {\bf 74} (2006) 023510.
\newblock \doi{10.1103/PhysRevD.74.023510}.

\bibitem[{Arcadi and Ullio(2011)}]{Arcadi:2011ev}
Arcadi G, Ullio P.
\newblock {Accurate estimate of the relic density and the kinetic decoupling in non-thermal dark matter models}.
\newblock {\em Phys. Rev. D\/} {\bf 84} (2011) 043520.
\newblock \doi{10.1103/PhysRevD.84.043520}.

\bibitem[{Gondolo and Gelmini(1991)}]{Gondolo:1990dk}
Gondolo P, Gelmini G.
\newblock {Cosmic abundances of stable particles: Improved analysis}.
\newblock {\em Nucl. Phys. B\/} {\bf 360} (1991) 145--179.
\newblock \doi{10.1016/0550-3213(91)90438-4}.

\bibitem[{Belanger et~al.(2007)Belanger, Boudjema, Pukhov, and Semenov}]{Belanger:2006is}
Belanger G, Boudjema F, Pukhov A, Semenov A.
\newblock {MicrOMEGAs $2.0$: A Program to calculate the relic density of dark matter in a generic model}.
\newblock {\em Comput. Phys. Commun.\/} {\bf 176} (2007) 367--382.
\newblock \doi{10.1016/j.cpc.2006.11.008}.

\bibitem[{Belanger et~al.(2009)Belanger, Boudjema, Pukhov, and Semenov}]{Belanger:2008sj}
Belanger G, Boudjema F, Pukhov A, Semenov A.
\newblock {Dark matter direct detection rate in a generic model with micrOMEGAs $2.2$}.
\newblock {\em Comput. Phys. Commun.\/} {\bf 180} (2009) 747--767.
\newblock \doi{10.1016/j.cpc.2008.11.019}.

\bibitem[{Gondolo et~al.(2004)Gondolo, Edsjo, Ullio, Bergstrom, Schelke, and Baltz}]{Gondolo:2004sc}
Gondolo P, Edsjo J, Ullio P, Bergstrom L, Schelke M, Baltz EA.
\newblock {DarkSUSY: Computing supersymmetric dark matter properties numerically}.
\newblock {\em JCAP\/} {\bf 0407} (2004) 008.
\newblock \doi{10.1088/1475-7516/2004/07/008}.

\bibitem[{Bringmann et~al.(2018)Bringmann, Edsj\"o, Gondolo, Ullio, and Bergstr\"om}]{Bringmann:2018lay}
Bringmann T, Edsj\"o J, Gondolo P, Ullio P, Bergstr\"om L.
\newblock {DarkSUSY 6 : An Advanced Tool to Compute Dark Matter Properties Numerically}.
\newblock {\em JCAP\/} {\bf 07} (2018) 033.
\newblock \doi{10.1088/1475-7516/2018/07/033}.

\bibitem[{Kawasaki et~al.(2000)Kawasaki, Kohri, and Sugiyama}]{Kawasaki:2000en}
Kawasaki M, Kohri K, Sugiyama N.
\newblock {MeV scale reheating temperature and thermalization of neutrino background}.
\newblock {\em Phys. Rev. D\/} {\bf 62} (2000) 023506.
\newblock \doi{10.1103/PhysRevD.62.023506}.

\bibitem[{Hannestad(2004)}]{Hannestad:2004px}
Hannestad S.
\newblock {What is the lowest possible reheating temperature?}
\newblock {\em Phys. Rev. D\/} {\bf 70} (2004) 043506.
\newblock \doi{10.1103/PhysRevD.70.043506}.

\bibitem[{Ichikawa et~al.(2005)Ichikawa, Kawasaki, and Takahashi}]{Ichikawa:2005vw}
Ichikawa K, Kawasaki M, Takahashi F.
\newblock {The Oscillation effects on thermalization of the neutrinos in the Universe with low reheating temperature}.
\newblock {\em Phys. Rev. D\/} {\bf 72} (2005) 043522.
\newblock \doi{10.1103/PhysRevD.72.043522}.

\bibitem[{De~Bernardis et~al.(2008)De~Bernardis, Pagano, and Melchiorri}]{DeBernardis:2008zz}
De~Bernardis F, Pagano L, Melchiorri A.
\newblock {New constraints on the reheating temperature of the universe after WMAP-5}.
\newblock {\em Astropart. Phys.\/} {\bf 30} (2008) 192--195.
\newblock \doi{10.1016/j.astropartphys.2008.09.005}.

\bibitem[{Arcadi et~al.(2020)Arcadi, Djouadi, and Raidal}]{Arcadi:2019lka}
Arcadi G, Djouadi A, Raidal M.
\newblock {Dark Matter through the Higgs portal}.
\newblock {\em Phys. Rept.\/} {\bf 842} (2020) 1--180.
\newblock \doi{10.1016/j.physrep.2019.11.003}.

\bibitem[{Arcadi et~al.(2019)Arcadi, Lebedev, Pokorski, and Toma}]{Arcadi:2019oxh}
Arcadi G, Lebedev O, Pokorski S, Toma T.
\newblock {Real Scalar Dark Matter: Relativistic Treatment}.
\newblock {\em JHEP\/} {\bf 08} (2019) 050.
\newblock \doi{10.1007/JHEP08(2019)050}.

\bibitem[{Aprile et~al.(2019)}]{XENON:2019gfn}
Aprile E, et~al.
\newblock {Light Dark Matter Search with Ionization Signals in XENON1T}.
\newblock {\em Phys. Rev. Lett.\/} {\bf 123} (2019) 251801.
\newblock \doi{10.1103/PhysRevLett.123.251801}.

\bibitem[{Aalbers et~al.(2023)}]{LZ:2022lsv}
Aalbers J, et~al.
\newblock {First Dark Matter Search Results from the LUX-ZEPLIN (LZ) Experiment}.
\newblock {\em Phys. Rev. Lett.\/} {\bf 131} (2023) 041002.
\newblock \doi{10.1103/PhysRevLett.131.041002}.

\bibitem[{Aad et~al.(2023)}]{ATLAS:2023tkt}
Aad G, et~al.
\newblock {Combination of searches for invisible decays of the Higgs boson using 139 fb\ensuremath{-}1 of proton-proton collision data at s=13 TeV collected with the ATLAS experiment}.
\newblock {\em Phys. Lett. B\/} {\bf 842} (2023) 137963.
\newblock \doi{10.1016/j.physletb.2023.137963}.

\bibitem[{Cepeda et~al.(2019)}]{Cepeda:2019klc}
Cepeda M, et~al.
\newblock {Report from Working Group 2}: {Higgs Physics at the HL-LHC and HE-LHC}.
\newblock {\em CERN Yellow Rep. Monogr.\/} {\bf 7} (2019) 221--584.
\newblock \doi{10.23731/CYRM-2019-007.221}.

\bibitem[{de~Blas et~al.(2020)}]{deBlas:2019rxi}
de~Blas J, et~al.
\newblock {Higgs Boson Studies at Future Particle Colliders}.
\newblock {\em JHEP\/} {\bf 01} (2020) 139.
\newblock \doi{10.1007/JHEP01(2020)139}.

\bibitem[{Moroi and Randall(2000)}]{Moroi:1999zb}
Moroi T, Randall L.
\newblock {Wino cold dark matter from anomaly mediated SUSY breaking}.
\newblock {\em Nucl. Phys. B\/} {\bf 570} (2000) 455--472.
\newblock \doi{10.1016/S0550-3213(99)00748-8}.

\bibitem[{Allahverdi et~al.(2013)Allahverdi, Dutta, Mohapatra, and Sinha}]{Allahverdi:2013tca}
Allahverdi R, Dutta B, Mohapatra RN, Sinha K.
\newblock {A Supersymmetric Model for Dark Matter and Baryogenesis Motivated by the Recent CDMS Result}.
\newblock {\em Phys. Rev. Lett.\/} {\bf 111} (2013) 051302.
\newblock \doi{10.1103/PhysRevLett.111.051302}.

\bibitem[{Acharya et~al.(2008)Acharya, Kumar, Bobkov, Kane, Shao, and Watson}]{Acharya:2008bk}
Acharya BS, Kumar P, Bobkov K, Kane G, Shao J, Watson S.
\newblock {Non-thermal Dark Matter and the Moduli Problem in String Frameworks}.
\newblock {\em JHEP\/} {\bf 06} (2008) 064.
\newblock \doi{10.1088/1126-6708/2008/06/064}.

\bibitem[{Acharya et~al.(2009)Acharya, Kane, Watson, and Kumar}]{Acharya:2009zt}
Acharya BS, Kane G, Watson S, Kumar P.
\newblock {A Non-thermal WIMP Miracle}.
\newblock {\em Phys. Rev. D\/} {\bf 80} (2009) 083529.
\newblock \doi{10.1103/PhysRevD.80.083529}.

\bibitem[{Moroi et~al.(2013)Moroi, Nagai, and Takimoto}]{Moroi:2013sla}
Moroi T, Nagai M, Takimoto M.
\newblock {Non-Thermal Production of Wino Dark Matter via the Decay of Long-Lived Particles}.
\newblock {\em JHEP\/} {\bf 07} (2013) 066.
\newblock \doi{10.1007/JHEP07(2013)066}.

\bibitem[{Aparicio et~al.(2016)Aparicio, Cicoli, Dutta, Muia, and Quevedo}]{Aparicio:2016qqb}
Aparicio L, Cicoli M, Dutta B, Muia F, Quevedo F.
\newblock {Light Higgsino Dark Matter from Non-thermal Cosmology}.
\newblock {\em JHEP\/} {\bf 11} (2016) 038.
\newblock \doi{10.1007/JHEP11(2016)038}.

\bibitem[{Chowdhury et~al.(2019)Chowdhury, Dudas, Dutra, and Mambrini}]{Chowdhury:2018tzw}
Chowdhury D, Dudas E, Dutra M, Mambrini Y.
\newblock {Moduli Portal Dark Matter}.
\newblock {\em Phys. Rev. D\/} {\bf 99} (2019) 095028.
\newblock \doi{10.1103/PhysRevD.99.095028}.

\bibitem[{Riajul~Haque et~al.(2023)Riajul~Haque, Kpatcha, Maity, and Mambrini}]{RiajulHaque:2023cqe}
Riajul~Haque M, Kpatcha E, Maity D, Mambrini Y.
\newblock {Primordial black hole reheating}.
\newblock {\em Phys. Rev. D\/} {\bf 108} (2023) 063523.
\newblock \doi{10.1103/PhysRevD.108.063523}.

\bibitem[{Hardy(2018)}]{Hardy:2018bph}
Hardy E.
\newblock {Higgs portal dark matter in non-standard cosmological histories}.
\newblock {\em JHEP\/} {\bf 06} (2018) 043.
\newblock \doi{10.1007/JHEP06(2018)043}.

\bibitem[{Pallis(2004)}]{Pallis:2004yy}
Pallis C.
\newblock {Massive particle decay and cold dark matter abundance}.
\newblock {\em Astropart. Phys.\/} {\bf 21} (2004) 689--702.
\newblock \doi{10.1016/j.astropartphys.2004.05.006}.

\bibitem[{Drees and Hajkarim(2018{\natexlab{a}})}]{Drees:2017iod}
Drees M, Hajkarim F.
\newblock {Dark Matter Production in an Early Matter Dominated Era}.
\newblock {\em JCAP\/} {\bf 02} (2018{\natexlab{a}}) 057.
\newblock \doi{10.1088/1475-7516/2018/02/057}.

\bibitem[{Drees and Hajkarim(2018{\natexlab{b}})}]{Drees:2018dsj}
Drees M, Hajkarim F.
\newblock {Neutralino Dark Matter in Scenarios with Early Matter Domination}.
\newblock {\em JHEP\/} {\bf 12} (2018{\natexlab{b}}) 042.
\newblock \doi{10.1007/JHEP12(2018)042}.

\bibitem[{Cheung et~al.(2011)Cheung, Elor, Hall, and Kumar}]{Cheung:2010gj}
Cheung C, Elor G, Hall LJ, Kumar P.
\newblock {Origins of Hidden Sector Dark Matter I: Cosmology}.
\newblock {\em JHEP\/} {\bf 03} (2011) 042.
\newblock \doi{10.1007/JHEP03(2011)042}.

\bibitem[{Figueroa and Tanin(2019)}]{Figueroa:2019paj}
Figueroa DG, Tanin EH.
\newblock {Ability of LIGO and LISA to probe the equation of state of the early Universe}.
\newblock {\em JCAP\/} {\bf 08} (2019) 011.
\newblock \doi{10.1088/1475-7516/2019/08/011}.

\bibitem[{Liu et~al.(2023)Liu, Chen, and Huang}]{Liu:2023pau}
Liu L, Chen ZC, Huang QG.
\newblock {Probing the equation of state of the early Universe with pulsar timing arrays}.
\newblock {\em JCAP\/} {\bf 11} (2023) 071.
\newblock \doi{10.1088/1475-7516/2023/11/071}.

\bibitem[{Charles et~al.(2016)}]{Fermi-LAT:2016afa}
Charles E, et~al.
\newblock {Sensitivity Projections for Dark Matter Searches with the Fermi Large Area Telescope}.
\newblock {\em Phys. Rept.\/} {\bf 636} (2016) 1--46.
\newblock \doi{10.1016/j.physrep.2016.05.001}.

\bibitem[{McDaniel et~al.(2024)McDaniel, Ajello, Karwin, Di~Mauro, Drlica-Wagner, and S\'anchez-Conde}]{McDaniel:2023bju}
McDaniel A, Ajello M, Karwin CM, Di~Mauro M, Drlica-Wagner A, S\'anchez-Conde MA.
\newblock {Legacy analysis of dark matter annihilation from the Milky~Way dwarf spheroidal galaxies with 14~years of Fermi-LAT data}.
\newblock {\em Phys. Rev. D\/} {\bf 109} (2024) 063024.
\newblock \doi{10.1103/PhysRevD.109.063024}.

\bibitem[{Galli et~al.(2009)Galli, Iocco, Bertone, and Melchiorri}]{Galli:2009zc}
Galli S, Iocco F, Bertone G, Melchiorri A.
\newblock {CMB constraints on Dark Matter models with large annihilation cross-section}.
\newblock {\em Phys. Rev. D\/} {\bf 80} (2009) 023505.
\newblock \doi{10.1103/PhysRevD.80.023505}.

\bibitem[{Ballesteros et~al.(2021)Ballesteros, Garcia, and Pierre}]{Ballesteros:2020adh}
Ballesteros G, Garcia MAG, Pierre M.
\newblock {How warm are non-thermal relics? Lyman-$\alpha$ bounds on out-of-equilibrium dark matter}.
\newblock {\em JCAP\/} {\bf 03} (2021) 101.
\newblock \doi{10.1088/1475-7516/2021/03/101}.

\end{thebibliography}
\end{document}